\documentclass[pra,onecolumn,epsf]{revtex4}
\usepackage{amsmath}
\usepackage{graphicx}
\usepackage[dvips]{color}
\usepackage{array}
\begin{document}

\title{Experimental observation of double-peak structure of coincidence spectra in ultra-fast Spontaneous Parametric Down Conversion}
\author{Dmitry A. Kalashnikov$^{1}$, Mikhail V. Fedorov$^{2}$, Leonid A. Krivitsky$^{1}$}
\email{Leonid_Krivitskiy@dsi.a-star.edu.sg; fedorovmv@gmail.com}
\affiliation{$^1$ Data Storage Institute, Agency for Science Technology and Research (A-STAR), 5 Engineering Drive I, 117608 Singapore\\
$^2$ A.M. Prokhorov General Physics Institute, Russian Academy of Sciences, 38 Vavilov st., Moscow, 119991, Russia} \vskip 24pt

\begin{abstract}
\begin{center}\parbox{14.5cm}
{The requirement on the symmetry of the biphoton wave function results in a double peak structure of the coincidence spectra of Spontaneous Parametric Down Conversion (SPDC) \cite{Scripta}. In this work we report the experimental observation of this effect by \textcolor{black}{carefully} tailoring parameters of the SPDC crystal, the ultra-fast pump laser, and the measurement setup. The results are shown to be relevant to quantum state engineering of ultra-fast polarization-, or frequency-entangled mixed states.}
\end{center}
\end{abstract}
 \maketitle \narrowtext
\vspace{-10mm}

\section{Introduction}
Spontaneous Parametric Down Conversion (SPDC) is a process in which pump photons decay in a nonlinear birefringent crystal into pairs of photons of lower frequencies, referred to as signal and idler photons (biphotons) \cite{DNK}, with the phasematching conditions being satisfied.  The SPDC provides a way for producing entangled states of light which are actively used in quantum optics community \cite{Nielsen&Chang}. The SPDC pumped by short laser pulses benefits the timing information of photon pairs and high repetition rate which can be useful in practical applications \cite{Sergienko, Kurt, Mosley, Branczyk}. At the same time, in the case of short pulses of the pump, the SPDC spectrum becomes wide enough for emitted photons to have a wide range of frequencies, which raises the task of measuring idler- and signal-photon spectra. A common technique to carry out such investigations is to perform a complete set of coincidence measurements with varying frequencies of both emitted photons \cite{Kurt, Keller, Grice, kim}. The coincidence spectrum of signal photons at a given frequency of the idler photons is usually characterized by a rather narrow curve, the width of which can be considered as a spectral measure of the correlation of photon pairs \cite{PRA}. However, as it was shown in \cite{Scripta}, in the case of short pump pulses and type-II phase matching under appropriate conditions, such coincidence spectrum can be more complicated and can have two well separated and well pronounced peaks. This effect can be clearly explained by the phase matching conditions and the differences between the refractive indices of the ordinary and extraordinary waves in birefringent crystals. However, in theory, the main reason of the peak-doubling effect is related to the symmetry requirements imposed on the biphoton wave functions. As photons are bosons, and if the SPDC biphoton state is pure, then its wave function must be symmetric with respect to the particle transpositions \cite{bosons}. If the wave function does not appear to be automatically symmetric (as in the case of the type-I phase matching \cite{PRA}), it has to be completed by the addition of terms with transposed particle variables in order to become symmetric. This is the way in which a double-peak has been found theoretically \cite{Scripta}. We confirm the existence of the double-peak structure by direct experimental measurements. In accordance with the theory, the double-peak structure exists only in the case of pulses shorter than 200 fs, and under specially arranged conditions of the experiment. Note that in 2007 a closely related experiment was performed \cite{Kurt}, \textcolor{black}{where it was shown that the two well defined decay paths lead to degradation of the visibility of polarization entanglement.} However the double peak structure was not directly observed in that work.

\section{Theory}
Consider the type-II collinear SPDC process pumped by short (femtosecond) laser pulses: let both the pump-pulse temporal envelope and its Fourier transform be approximated by Gaussian functions. The spectral features of the signal and idler photons are determined by their wave function, which will in turn depend on the photon frequencies $\omega_1$ and $\omega_2$ as well as the condition that it has to be symmetric with respect to the variable transposition $1\rightleftharpoons 2$. Such wave function can be presented in the form \cite{Scripta}
{\color{black}
\begin{eqnarray}
 %\nonumber
 \Psi(\omega_1,\omega_2)\propto \exp\left(-\frac{(\omega_1+\omega_2-\omega_0)^2\tau^2}{8\ln 2}\right)
 \label{symmWF}
 \times\left\{\left(1\atop 0\right)_1\left(0\atop 1\right)_2{\rm sinc}\left[\frac{L}{2}\Delta_{12}\right]
 +\left(1\atop 0\right)_2\left(0\atop 1\right)_1{\rm sinc}\left[\frac{L}{2}\Delta_{21}\right]\right\},
\end{eqnarray}
where $\omega_0$ is the central frequency of the pump, $\tau$ is the pump-pulse duration, and $L$ is the crystal length. The two-line columns correspond to different polarizations of photons, $\left(1\atop 0\right)$ for horizontal and $\left(0\atop 1\right)$ for vertical polarizations, and the indices 1 and 2 are associated with the numbers of frequency variables $\omega_1$ and $\omega_2$ of photons. $\Delta_{12}$ and $\Delta_{21}$ in Eq. (\ref{symmWF}) are the phase mismatches differing from each other by the transposition of $\omega_1$ and $\omega_2$}
\begin{equation}
 \label{Del12-Del21}
 \setlength{\extrarowheight}{0.3cm}
 \begin{matrix}
 \Delta_{12}(\omega_1,\omega_2)=k_p\left(\omega_1+\omega_2, \phi_0\right)-k_o(\omega_1)-k_e\left(\omega_2, \phi_0\right)
 =n_e\left(\omega_1+\omega_2,\,\phi_0\right)\frac{\omega_1+\omega_2}{c}
 -n_o(\omega_1)\frac{\omega_1}{c}-n_e\left(\omega_2,\phi_0\right)\frac{\omega_2}{c},\\
 \Delta_{21}(\omega_1,\omega_2)=k_p\left(\omega_1+\omega_2, \phi_0\right)-k_o(\omega_2)-k_e\left(\omega_1, \phi_0\right)
 =n_e\left(\omega_1+\omega_2,\,\phi_0\right)\frac{\omega_1+\omega_2}{c}
 -n_o(\omega_2)\frac{\omega_2}{c}-n_e\left(\omega_1,\phi_0\right)\frac{\omega_1}{c}.
 \end{matrix}
\end{equation}
Here $k_o,n_o$ and $k_e,n_e$ are wave numbers and refractive indices which correspond to ordinary (\textit{o}) and extraordinary (\textit{e}) waves in the crystal, and $\phi_0$ is the angle between the crystal optical axis and the direction of photon propagation. {\color{black}The angle $\phi_0$ is determined by the condition of zero mismatches at frequencies equal to half of the pump central frequency for both terms in Eqs. (\ref{symmWF}) and (\ref{Del12-Del21}): $\Delta_{12}(\omega_0/2,\omega_0/2)=\Delta_{21}(\omega_0/2,\omega_0/2)=0$. At any other coinciding frequencies $\omega_1=\omega_2\neq\omega_0/2$ mismatches $\Delta_{12}$ and $\Delta_{21}$ (\ref{Del12-Del21}) are equal to each other too, but they differ from zero,  $\Delta_{12}(\omega,\omega)=\Delta_{21}(\omega,\omega)\neq 0$ at $\omega\neq\omega_0/2$. But if the frequencies  $\omega_1$ and $\omega_2$ do not coincide with each other, the mismatches $\Delta_{12}$ and $\Delta_{21}$ (\ref{Del12-Del21}) do not coincide with each other also: $\Delta_{12}\neq\Delta_{21}$ at
$\omega_1\neq\omega_2$}.

The appearance of two terms in Eq. (\ref{symmWF}) is related to the obligatory requirement that the biphoton wave function must be symmetric with respect to the transposition of photon variables $1\rightleftharpoons 2$.  An alternative explanation (not contradicting to the first one) is based on the observation that in the type-II SPDC process there are two possibilities to measure coincidence spectra: the fixed frequency $\omega_2$ can be attributed either to the extraordinary (vertically polarized)  or to the ordinary (horizontally polarized) waves. These two cases exactly correspond  to the first and second terms in Eq. (\ref{symmWF}) and to two different mismatches $\Delta_{12}$ and $\Delta_{21}$ (\ref{Del12-Del21}). Note that the assumed correspondence between polarizations of ordinary, and extraordinary waves in the crystal is determined by the conditions of the experiment described in the following section.

Coincidence spectra of emitted photons are determined by the conditional probability densities $\left.dw(\omega_1)/d\omega_1\right|_{\omega_2}$, where $\omega_1$ is varying and $\omega_2=const$.  There are two  ways of performing coincidence measurements: after being split by a 50/50 non-polarizing beamsplitter the down-converted photons are either filtered by polarizers and frequency filters in front of the detectors or just by the frequency filters.  In the first case the spectral conditional probabilities become simultaneously conditional with respect to the polarization and are given by squared absolute values of the functions accompanying polarization columns in Eq. (\ref{symmWF}):
\begin{equation}
\left.\frac{dw_{\color{black}H}(\omega_1)}{d\omega_1}\right|_{{\color{black}V},\,\omega_2}\propto\exp\left(-\frac{(\omega_1+\omega_2-\omega_0)^2\tau^2}{4\ln 2}\right)
 \label{cond-12}
 {\rm sinc}^2\left[\frac{L}{2}\Delta_{12}(\omega_1,\omega_2)\right],
\end{equation}
and
\begin{equation}
\left.\frac{dw_V(\omega_1)}{d\omega_1}\right|_{H,\,\omega_2}\propto\exp\left(-\frac{(\omega_1+\omega_2-\omega_0)^2\tau^2}{4\ln 2}\right)
 \label{cond-21}
 {\rm sinc}^2\left[\frac{L}{2}\Delta_{21}(\omega_1,\omega_2)\right].
\end{equation}
In the second case, when polarizers are not used \textcolor{black}{at all, one has to use the wave function (\ref{symmWF}) for constructing the
density matrix $\rho=\Psi\otimes\Psi^\dag$ depending on both polarization and frequency variables. The trace of this density matrix with
respect the polarization variables yields the density matrix of the mixed state depending on four frequency variables 
${\bar\rho}(\omega_1,\omega_2;\omega_1^\prime,\omega_2^\prime)$.
Diagonal elements of ${\bar\rho}$ determine two-frequency probability density
\begin{equation}
 \label{2-fr-prob}
 \frac{dw(\omega_1,\omega_2)}{d\omega_1d\omega_2}={{\bar\rho}(\omega_1,\omega_2;\omega_1,\omega_2)}
 \propto \exp\left(-\frac{(\omega_1+\omega_2-\omega_0)^2\tau^2}{4\ln 2}\right)
 \left\{{\rm sinc}^2\left[\frac{L}{2}\Delta_{12}(\omega_1,\omega_2)\right]+
 {\rm sinc}^2\left[\frac{L}{2}\Delta_{21}(\omega_1,\omega_2)\right]\right\},
\end{equation}
}
as well as the conditional (coincidence) probability density
\begin{eqnarray}
 \nonumber
 \frac{dw^{(c)}(\omega_1)}{d\omega_1}=\left.\frac{dw(\omega_1,\omega_2)}{d\omega_1d\omega_2}
 \right|_{\omega_2=const}
 \propto\exp\left(-\frac{(\omega_1+\omega_2-\omega_0)^2\tau^2}{4\ln 2}\right)\\
 \label{cond-tot}
  \times\left.\left\{{\rm sinc}^2\left[\frac{L}{2}\Delta_{12}(\omega_1,\omega_2)\right]+
 {\rm sinc}^2\left[\frac{L}{2}\Delta_{21}(\omega_1,\omega_2)\right]
  \right\}\right|_{\omega_2=const}.
\end{eqnarray}
As seen from Eq. (\ref{cond-tot}), the conditional probability density found for the mixed state ${\bar\rho}$ coincides with the sum of two probability densities (\ref{cond-12}) and (\ref{cond-21}) arising in the original pure state (\ref{symmWF}), (\ref{Del12-Del21}).  There are two ways of obtaining the same polarization-non-selective coincidence spectrum. One method consists in using only frequency filters with no polarizers. By keeping a given photon frequency $\omega_2$ in one channel after the beamsplitter and by scanning the filter in the other channel one obtains the spectrum $dw^{(c)}(\omega_1)/d\omega_1$ (\ref{cond-tot}), which is determined by the frequency-dependent mixed state with the density matrix ${\bar\rho}$. An alternative method consists of making two coincidence measurements with the polarizer added to the frequency filter in the given-frequency channel, and the polarizer oriented vertically in one measurement and horizontally in another one. Each of these two measurements will give rise to a single-peak spectrum (\ref{cond-12}) or (\ref{cond-21}), but their sum will coincide with the double-peak structure determined by Eq. (\ref{cond-tot}).

\begin{figure}[h]
\centering
		\includegraphics[width=0.3\textwidth]{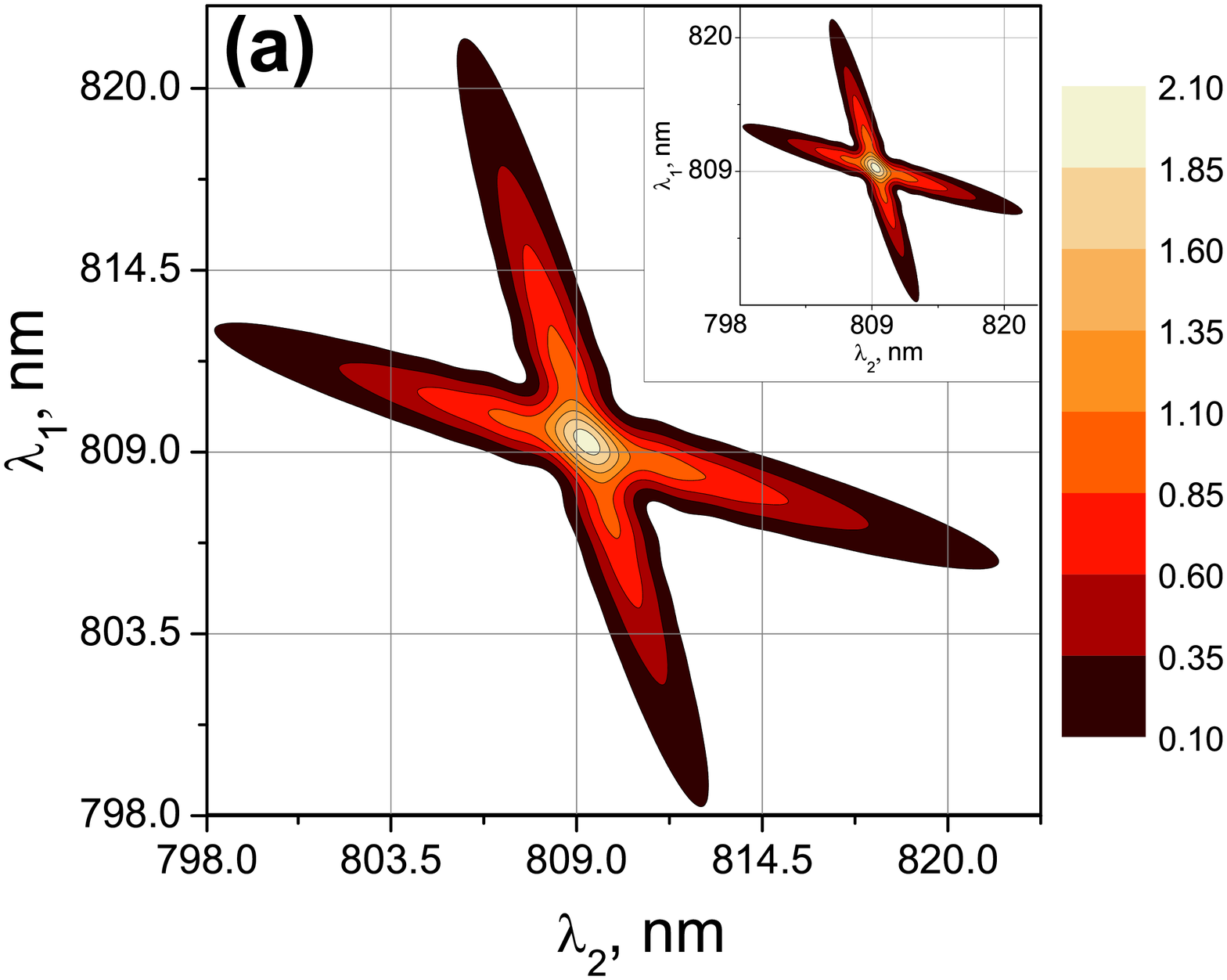}
		\includegraphics[width=0.3\textwidth]{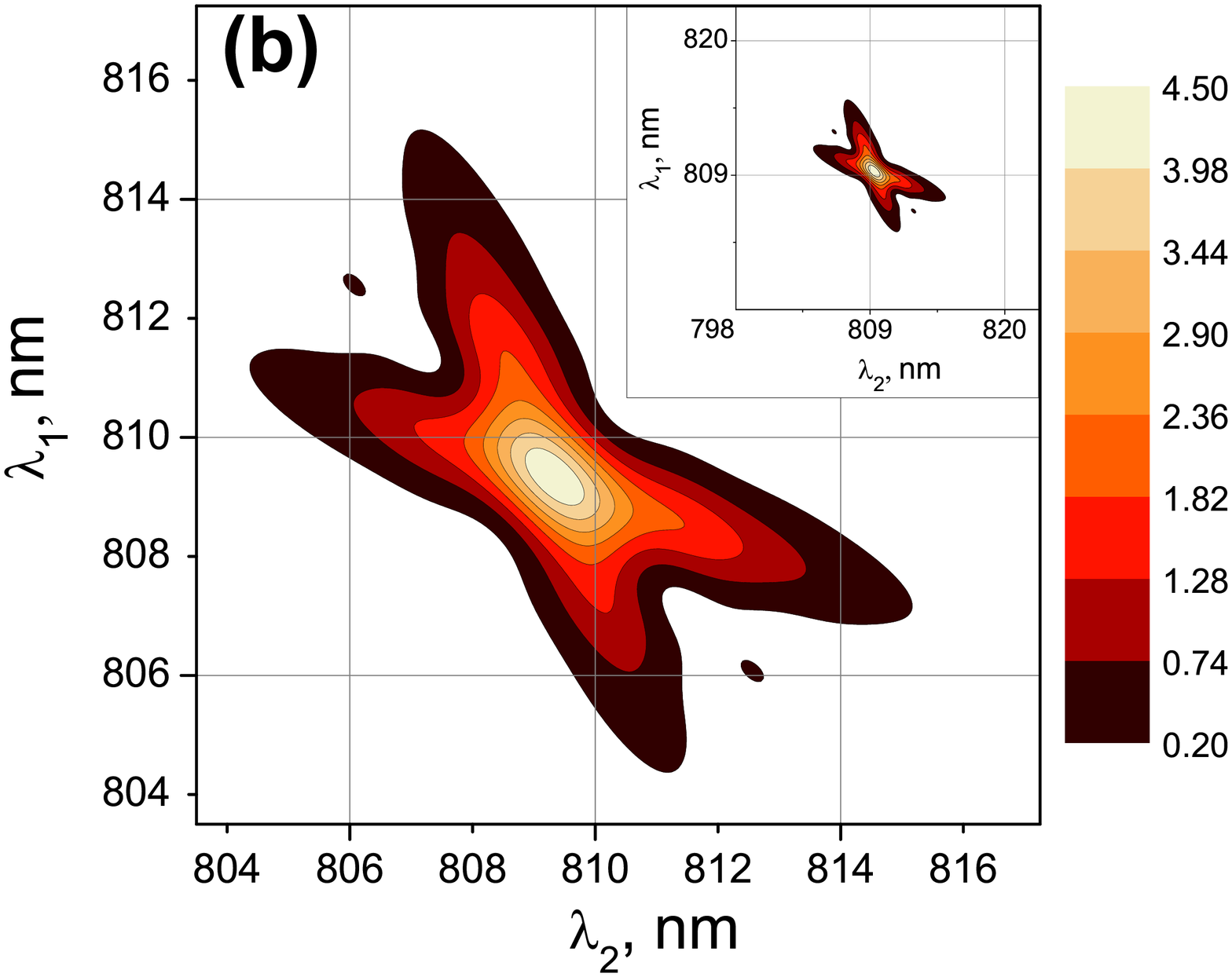}
		\includegraphics[width=0.3\textwidth]{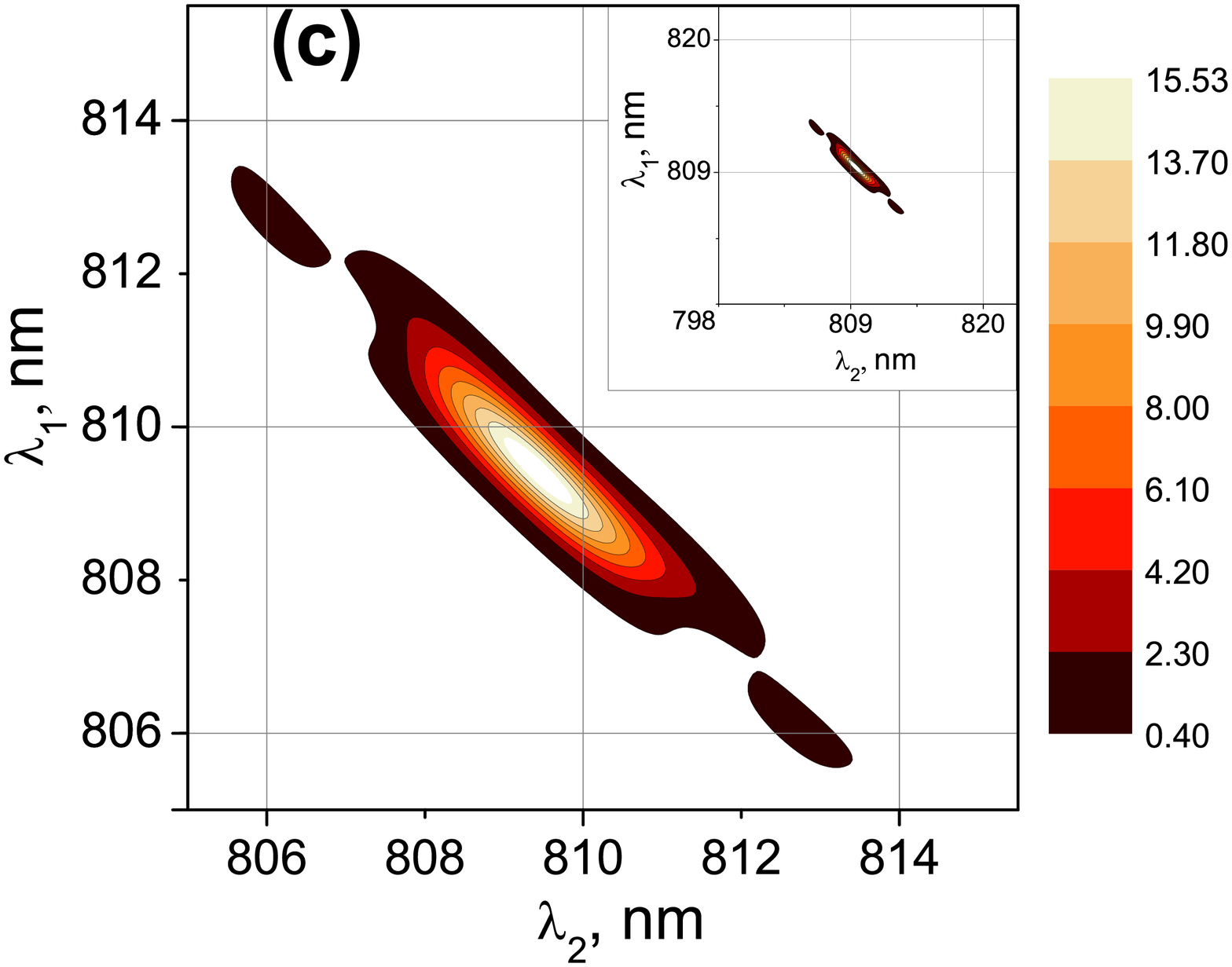}
	\caption{Calculation of the spectral conditional probability {\color{black}density} for type-II $\beta-$Barium borate (BBO) crystal of the length $L=5$ mm, optical axis at $\phi_{0}=41.4625^{\circ}$,  pump central wavelength $\lambda_{0}=404.7$ nm, duration of the pump pulse (a) $\tau=110$ fs, (b) $\tau=250$ fs, (c) $\tau=1000$ fs. \textcolor{black}{The insets in (a), (b), and (c) show the corresponding dependencies plotted on the same scale as (a).} The energy of the pump pulse is considered to be the same for the three cases , i.e. $\tau E_{0}^{2}=const$, where $E_{0}$ is the amplitude of the pump field.
\label{butterfly}}	
\end{figure}

Fig.\ref{butterfly} shows calculation of the spectral conditional probability {\color{black}density} according to Eq.(\ref{cond-tot}) for different pump-pulse durations. All frequency dependencies are recalculated to depend on the wavelengths $\lambda_{0,1,2}=2\pi c/\omega_{0,1,2}$ (for convenience of comparisons with the experimental data). The parameters of the calculations correspond to the ones used in the experiment (see below). For relatively short pump-pulses of $\tau=110$ fs (see Fig.1a) the spectral conditional probability  {\color{black}density} has a distinctive "butterfly" shape. Such a shape implies that for a fixed wavelength of the signal(idler) photon, slightly detuned from the degeneracy, the spectral conditional probability exhibits two distinctive maxima. Each maximum, corresponds to a different polarization component of the two-photon state.

The effect of peak doubling depends crucially on the duration of the pump-pulse. The corresponding dependencies of the spectral conditional probability  {\color{black}density} for the case of $\tau=250$ fs, and $\tau=1000$ fs are shown in Fig.\ref{butterfly}b and Fig.\ref{butterfly}c, respectively. In the case of $\tau=1000$ fs the  {\color{black}dependence} has only a single maximum at a fixed wavelength of the signal (idler) photon, and the effect of peak doubling vanishes. Disappearance of the peak doubling in the case of long pulses has a simple qualitative explanation. For states with the  wave function or density matrix having the form of a product of two parts, each depending on two frequencies, widths of conditional spectral distributions are determined mainly by the narrower part of the product. In the scheme under consideration, these two terms are the pump spectral intensity and the sum of two ${\rm sinc^2}$-functions in Eq. (\ref{cond-tot}). In the case of short pulses\textcolor{black}{,} the pump spectrum is wide, and the spectral structure of conditional probabilities is determined mainly by the sum of ${\rm sinc^2}$-functions, which gives rise to two peaks. In the opposite case of long pulses, the spectrum of the pump becomes very narrow, and the conditional spectra of idler-signal photons are determined exclusively by the pump. As the pump spectral intensity depends only on the sum of frequencies $\omega_1+\omega_2$ centered at $\omega_0$, it is clear that at any given value of $\omega_2$, the conditional spectrum {\color{black} $dw^{(c)}(\omega_1)/d\omega_1$ (\ref{cond-tot})} has only one peak, and the locations of $\omega_1$ and $\omega_2$ are symmetric with respect to $\omega_0$. This symmetry explains why the narrow region of large conditional probabilities in Fig.\ref{butterfly}c is concentrated around the line tilted at $45^\circ$ to both the horizontal and vertical axes.

\section{Experiment}

\begin{figure}[h]
\centering
		\includegraphics[width=0.8\textwidth]{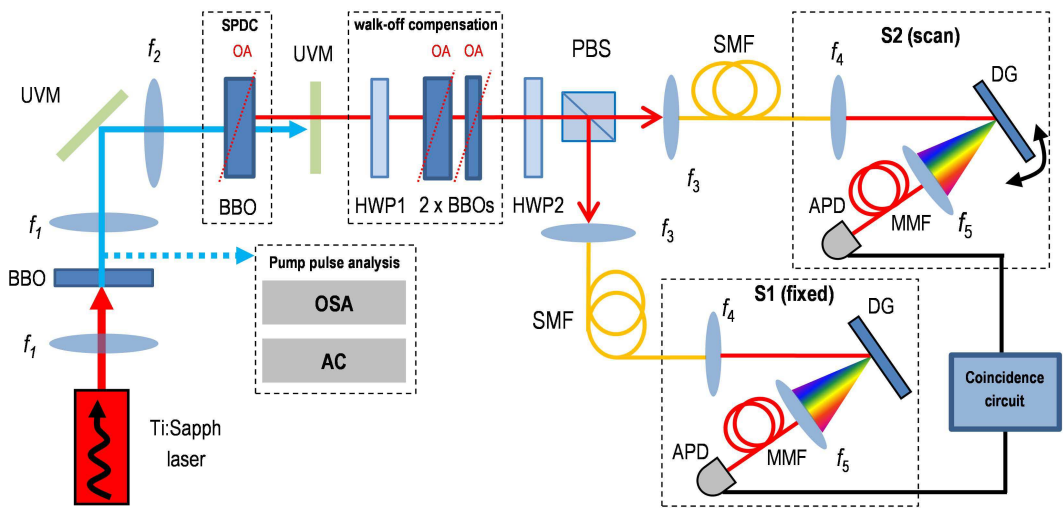}
	\caption{Experimental setup. The beam of a mode-locked Ti:Sapphire laser is frequency doubled, and it is then focused into a type-II BBO crystal, where SPDC occurs. The second harmonic (SH) is characterized by an optical spectrum analyzer (OSA), and an autocorrelator (AC). UV-Mirror (UVM) filters the SPDC from the pump. A half-wave plate (HWP1) and two BBO crystals with parallel optical axes (OA) compensate for the spatial walk-off. The SPDC passes through a half-wave plate (HWP2), and divided by a polarization beam splitter (PBS). It is coupled into two single-mode fibers (SMF), connected to two spectrometers (S1, S2). Each spectrometer consists of collimating and focusing lenses, and a diffraction grating (DG). Outputs of both spectrometers are directed to avalanche photodiodes (APDs) via multi-mode optical fibers (MMFs). APD outputs are \textcolor{black}{sent} to a coincidence circuit.
	\label{setup}}
\end{figure}

The experimental setup is shown in Fig.\ref{setup}. The beam of a mode-locked Ti:Sapphire laser (Spectra Physics, Mai-Tai) with a central wavelength of 809.4 nm, pulse duration 80 fs at full width half maximum (FWHM), repetition rate 80 MHz, and output power 1.1 W is frequency doubled in a 0.3 mm long $\beta$-Barium borate crystal (BBO), placed between two confocal lenses with \textit{f$_{1}$}=50 mm. The second harmonic (SH) beam is filtered from the fundamental beam by low group velocity dispersion dichroic mirrors (not shown), and characterized by an autocorrelator (AC, Femtochrome, FR-103XL), and an optical spectrum analyzer (OSA, Yokogawa, AQ-6315B). The SH beam has a central wavelength 404.7 nm, spectral bandwidth 3.5 nm (FWHM), duration 110 fs (FWHM), and power 200 mW. It is focused by a lens with \textit{f$_{2}$}=500 mm into a 5 mm long BBO crystal cut for type-II (cut angle $\phi_0=41.45^{\circ}$; \textcolor{black}{accuracy $<0.5^{\circ}$}) collinear frequency-degenerate SPDC at 810 nm. The SH beam is vertically polarized (extraordinary wave in the crystal) and its waist inside the crystal is 200 $\mu$m. After the SPDC crystal the SH beam is reflected by a UV-mirror (UVM) whilst the SPDC fluently passes through it. In order to compensate for the spatial walk-off in the SPDC crystal, a half-wave plate orientated at $45^\circ$ (HWP1) and two BBO crystals 2 and 0.5 mm long (total length is equal to the  half of the length of the SPDC crystal) with their optical axes parallel to the one of the SPDC crystal, are introduced \cite{Kwiat}. Then, the SPDC passes through another half-wave plate (HWP2) followed by a polarization beam splitter (PBS). Such a configuration allows for the routing of different polarization components of the SPDC to different output ports of the PBS by rotating the HWP2. Next, the SPDC in each output port of the PBS is coupled into a single mode optical fiber (SMF) by aspheric lens with \textit{f$_{3}$}=8 mm, placed at the distance 600 mm from the SPDC crystal \cite{Kurtsiefer}. Each SMF is connected to the input of a home-made spectrometer (S1, S2), which consists of an achromatic collimation lens with \textit{f$_{4}$}=60 mm, a diffraction grating (Thorlabs, GR-1208) mounted on a motorized rotation stage, and a collection lens with \textit{f$_{5}$}=175 mm, which focuses the light into a multi-mode optical fiber (MMF) with the core diameter of 50 $\mu$m. The MMF is connected to an avalanche photodiode (APD, Perkin-Elmer, AQRH-14FC) with quantum efficiency 60\%. Outputs of the APDs from the two spectrometers are addressed to a coincidence circuit (Ortec, Time to Amplitude Converter, Model 556) with the time window set to 7 ns. Each spectrometer is tested with a fiber coupled broadband light source, and yields 0.15 nm spectral resolution and 40\% transmission. From the theoretical calculations\textcolor{black}{, using Eq. (\ref{cond-tot}),} it is found that slight changes in the orientation of the crystal axis (on the order of $0.0005^\circ$) results in significant changes of the observed double-peak structure of the coincidence spectra. To circumvent this problem all the presented data has been obtained in a single experimental run without realignment of the setup.

\section{Results and discussion}

\begin{figure}
\centering
		\includegraphics[width=0.35\textwidth]{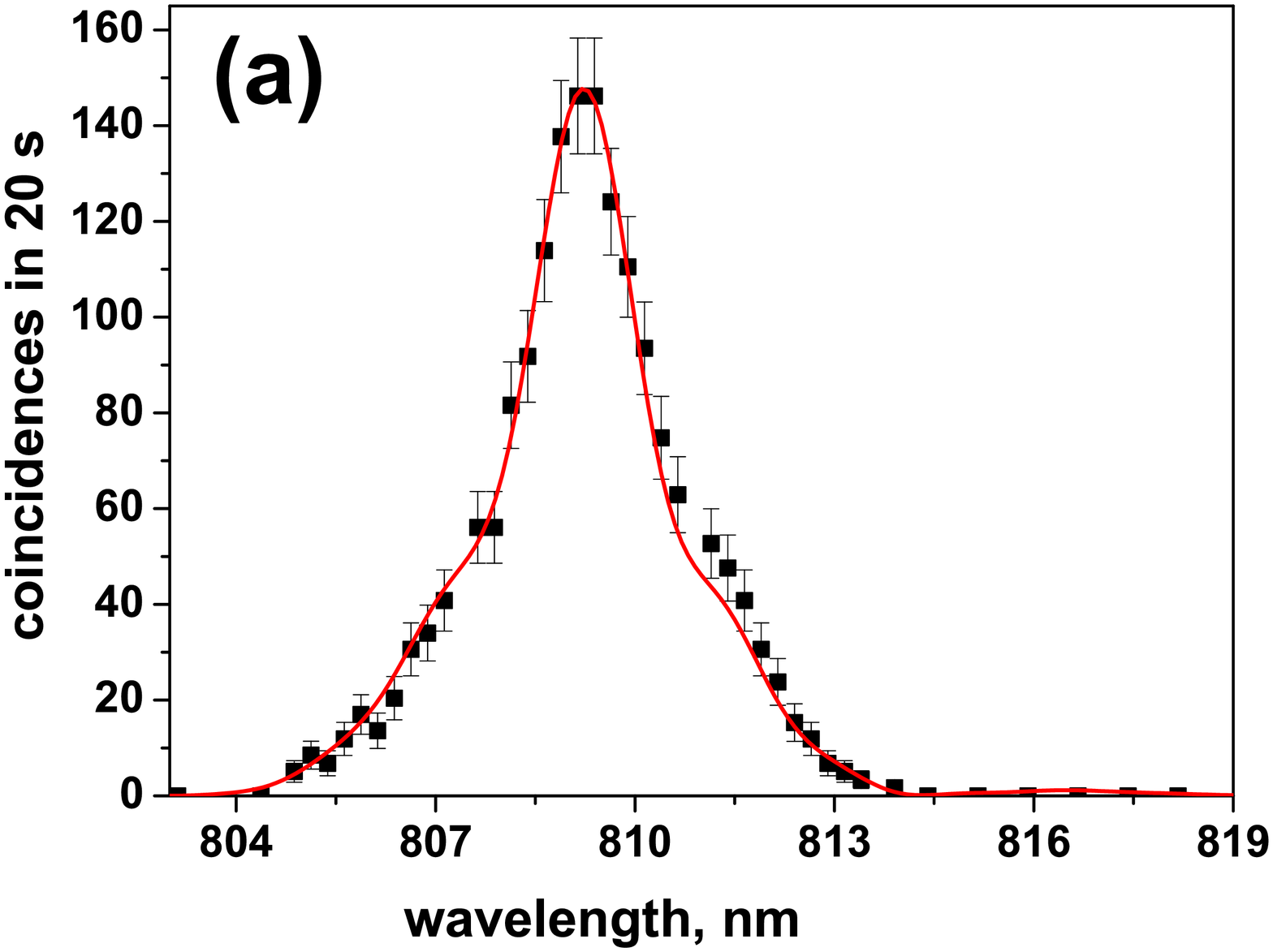}
		\includegraphics[width=0.35\textwidth]{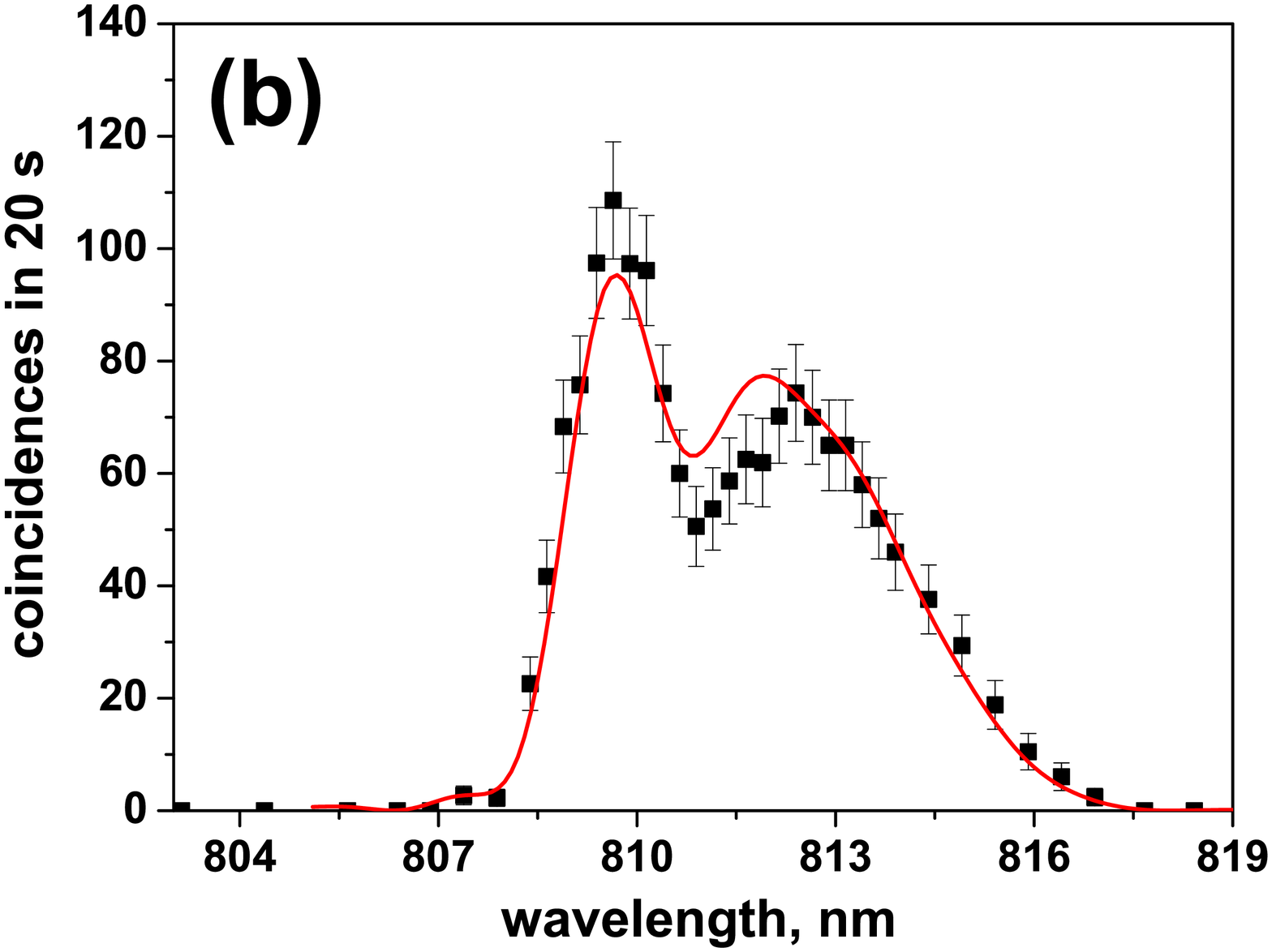}\\
		\includegraphics[width=0.35\textwidth]{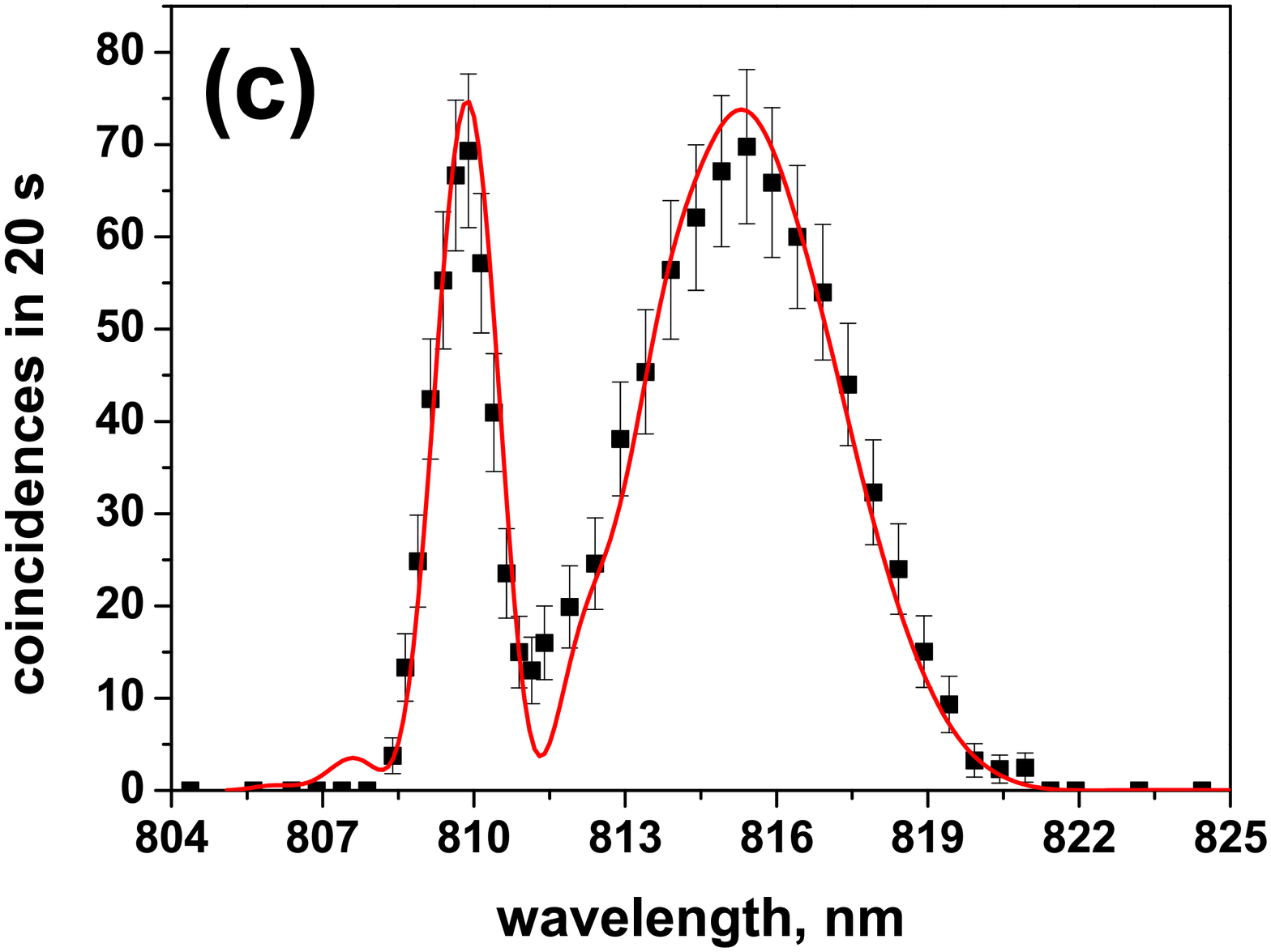}
		\includegraphics[width=0.35\textwidth]{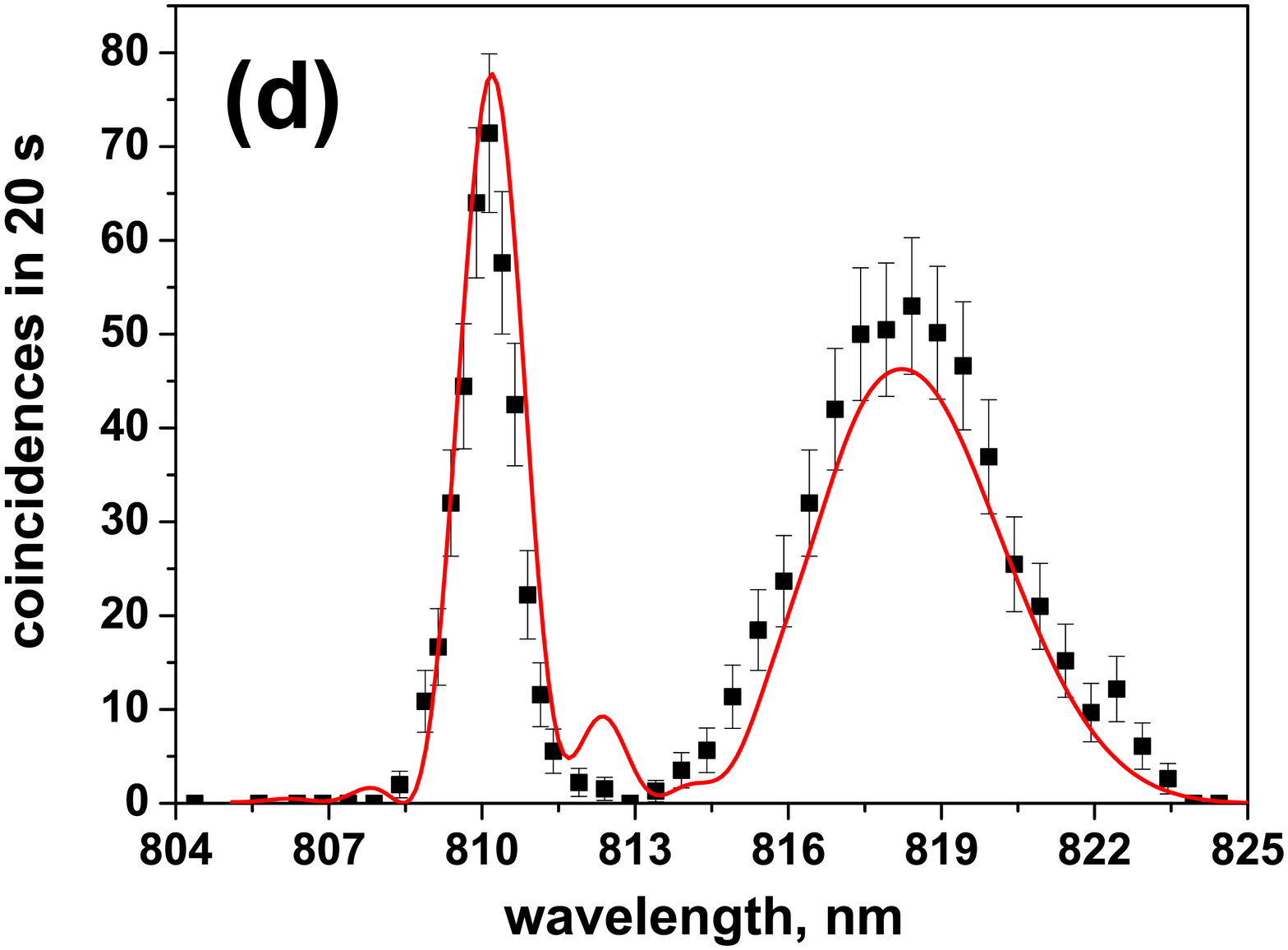}
	\caption{Experimental results. Spectral dependencies of the coincidences at different fixed wavelengths of an idler photon: (a) 809.4 nm, (b) 808.4 nm (c) 807.3 nm (d) 806.3 nm. Black points are the experimental data, error bars are the standard deviations, red solid lines are the theoretical calculations.\label{results}}
	\end{figure}

The wavelength (idler), selected by one of the spectrometers (S1), was fixed. In another spectrometer (S2) the wavelength (signal) was scanned and the spectral dependence of the coincidences was measured for a given polarization component. Then, HWP2, initially set at $0^\circ$, was rotated by $45^\circ$ and the above mentioned procedure is carried out for another polarization component. The overall coincidence spectrum is obtained by summation of the measurements from the two orientations of HWP2. Spectral dependencies of the coincidences are tested at four different wavelengths of idler photons selected by the spectrometer S1: 809.4 nm, 808.4 nm, 807.3 nm, and 806.3 nm. The corresponding results are shown in Fig.~\ref{results}.

The experimental data (black squares) demonstrate a fair agreement with the theory (red solid lines). With the wavelength of the idler photon detuned from the degeneracy, the spectral conditional probability exhibits two distinctive maxima. The separation between the two maxima grows with the detuning of the idler photon wavelength, in accordance with the theoretically calculated spectral conditional probability, using (9). The only free parameter for fitting the theoretical calculations to the experimental data was the orientation of the optical axis of the crystal\textcolor{black}{, which was fixed to be the same for all the experimental plots}. The optimal value was found to be $\phi_0=41.4625^{\circ}\pm0.0001^{\circ}$. \textcolor{black}{The values of the coefficient of determination (COD($R^{2}$)) for the obtained fits are 0.991 (a), 0.933 (b), 0.956 (c), 0.889 (d).
Slight discrepancy between the experimental results and the theory is most likely caused by drift of the orientation of the optical axis during the experiment due to mechanical and thermal instabilities. Note that a small bump between two peaks in Fig.~\ref{results}d should not be misinterpreted as an additional third peak. Most probably, its appearance is related to specific features of the sinc-function, which does not fall monotonously at far wings of its main maximum.}

The curve of Fig. 1a corresponds to the case when the idler wavelength $\lambda_2$ is equal to the coinciding central wavelengths of both peaks. Summed together, they produce a single-peak curve of a rather unusual shape, with the narrower peak seeming to sit on top of the wider one. This combined peak is approximately twice as high as the two constituent individual peaks from which this curve is formed. This confirms that the observed results correspond to the mixed rather than pure polarization state of two photons, because in the case of a pure polarization state, coherent summation of individual peaks with coinciding central wavelengths results in a four fold increase of the maximal probability compared to that occurring in individual peaks. This is not observed. Hence, the experimental results support the theoretical description in terms of the mixed polarization state.

\section{Conclusion}

The presented work shows experimental observation of the earlier predicted effect of  peak doubling in the coincidence  spectra of type-II SPDC and ultra-fast pump pulses  \textcolor{black}{\cite{Scripta}}. The effect occurs only if the pulse duration of the pump is short enough to provide a pump spectrum broader than the widest of two peaks. In the case of longer pulses, only one of two peaks survives (Fig.\ref{butterfly}c). The experiment has confirmed that the state which is characterized by frequency distributions of photons averaged over polarizations is mixed (the peak in Fig. \ref{results}a is only two times higher than each of the two terms contributing to its formation in Eq. (\ref{cond-tot})).  The peak-doubling effect is a direct consequence of the symmetry requirement of the biphoton wave functions resulting in the appearance of two terms in expression (\ref{symmWF}) for the biphoton polarization-frequency wave function. \textcolor{black}{The effect leads to degradation of quality of polarization entanglement \cite{Kurt}, and therefore it should be carefully considered in engineering of entangled quantum key distribution systems exploiting ultra-fast polarization entangled states \cite{EQKD}.}

\begin{acknowledgments}
We would like to thank Yvonne Gao for help at the initial stage of the experiment. The work was supported by A-STAR Investigatorship grant. MVF acknowledges the support of RFBR, grant 11-02-01043.
\end{acknowledgments}

\end{document}